%% file: alp_arxiv_rev.tex
\documentclass[prd,twocolumn,noshowpacs,linenumbers,groupedaddress,superscriptaddress,amsmath,amssymb]{revtex4}
\usepackage{graphicx}% Include figure files
\usepackage{dcolumn}% Align table columns on decimal point
\usepackage{bm}% bold math
\usepackage{amssymb}
\usepackage{enumerate}
\usepackage{multirow} 
\usepackage[symbol]{footmisc}

\usepackage{float}
\usepackage{lineno}
\newcommand\g{\gamma}
\newcommand\gagg{g_{a\g\g}}

\newcommand\ma{m_{a}}

\newcommand\agg{a\rightarrow \gamma \gamma}

\newcommand\kspio{K^0_S \to \pi^0 \pi^0}
\newcommand\klpio{K^0_L \to 3\pi^0}

\def\address{\@ifstar{\address@star}%
  {\@ifnextchar[{\address@optarg}{\address@noptarg}}}

\begin{document}
\title{ 
\begin{center}
 {\Large EUROPEAN LABORATORY FOR PARTICLE PHYSICS}
\end{center}
%\begin{frontmatter}
\vskip0.3cm
\hspace{-3.0cm}{\rightline{\rm  CERN-EP-2020-068}}
\vskip.7cm
Search for Axionlike and Scalar Particles with the NA64 Experiment}
%\collab{NOMAD Collaboration}
\input author_list.tex
%\date{June 17, 2009}% It is always \today, today,

%\clearpage
\begin{abstract}
\begin{center}
\vskip 0.2cm
{This publication is dedicated to the memory of our colleague Danila Tlisov.}
\end{center}
We carried out a model-independent search for light  scalar ($s$) and 
pseudoscalar axionlike  ($a$)  particles that couple to two photons
 by using the  high-energy  CERN SPS H4 electron beam.   
  The new particles, if they exist, could be 
produced through the Primakoff effect in interactions of hard bremsstrahlung photons generated by 100 GeV electrons in the NA64 active dump with virtual photons
provided by the nuclei of the dump.  The $a(s)$ would  penetrate the downstream HCAL module,  serving  as a shield,  and would be observed  either 
through their $a(s)\to \g\g$ decay in the rest of the HCAL detector, or as events with a large missing energy if the $a(s)$ decays downstream of the HCAL.
This method allows for the probing of the $a(s)$ parameter space, including those from generic axion models, inaccessible to previous experiments.\ 
 No evidence of such processes has been found from the analysis of the data corresponding to $2.84\times10^{11}$ electrons on target,
 allowing us to set new limits  on the $a(s)\gamma\gamma$-coupling strength  for $a(s)$ masses below 55 MeV.
\end{abstract}

%\pacs{14.80.-j, 12.60.-i, 13.20.Cz, 13.35.Hb}% PACS, the Physics and Astronomy
\maketitle
%\section{Introduction}
%\label{sec:intro}
\begin{figure*}[tbh!!]
\includegraphics[width=.5\textwidth]{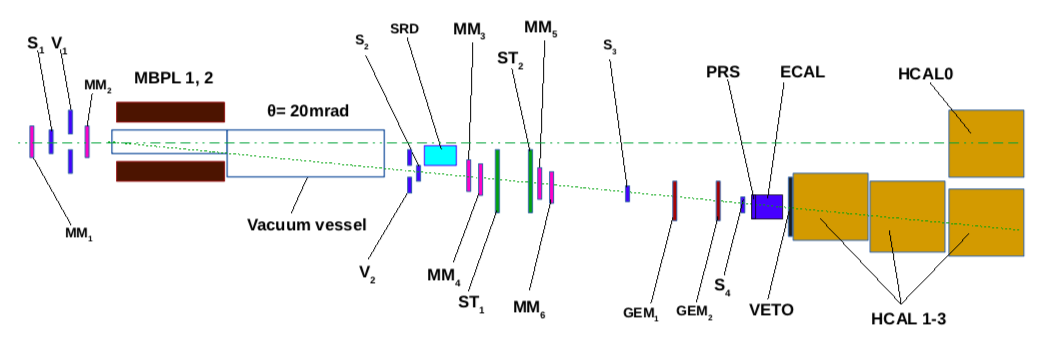}%
\includegraphics[width=0.3\textwidth]{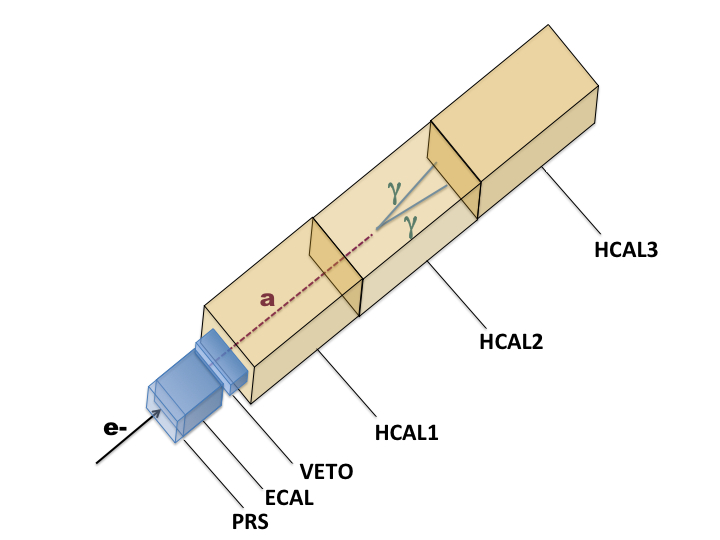}
\vskip-0.cm{\caption{The left panel illustrates schematic view of the setup to search for the $\agg$  decays of the $a$s 
produced in the reaction chain $e^-Z \to e^-Z \g; \g Z\to a Z$ induced by  100 GeV $e^-$s in the active ECAL dump. The right panel shows an example of  the  $\agg$  decay  
in the HCAL2 module. \label{setup}}}
\end{figure*} 
Neutral spin-zero scalar ($s$) or pseudoscalar ($a$)  massive  particles  are  predicted  in many extensions of the standard model (SM). 
The most popular light pseudoscalar, the axion, postulated  in  \cite{ww}
to provide a solution to the "strong CP" problem, emerges as a consequence of the 
breaking of the Peccei-Quinn (PQ) symmetry \cite{pq}.\ It is now believed  that the generic
axion  has a mass, perhaps  much smaller than $m_a \sim O(100)$ keV, which was originally 
expected \cite{dfsz, ksvz}.
The axionlike particles (ALPs), which are   pseudo-Nambu-Goldstone bosons,  
arise in models containing a spontaneously broken PQ symmetry, see, e.g., \cite{review, pdg}, 
with arbitrary masses and small couplings,  making them  natural candidates for the
mediator of interactions between dark and visible sectors or as candidate for dark matter (DM) themselves. 
ALPs could also provide a solution to  both the electron \cite{g-2e} and muon \cite{g-2mu}  $g-2$ anomalies \cite{marci}. 
This has motivated worldwide theoretical and experimental efforts towards dark forces and other portals between the visible and dark sectors,  
see, e.~g., ~\cite{Essig:2013lka, report1, report2, pbc-bsm, pbc, berlin,Feng:2018noy,Dobrich:2015jyk,Bauer:2018uxu,jaec,gkm, dvk}.
\par The $a-\g\g$  interaction is given by the Lagrangian 
\begin{equation}
L_{int}= - \frac{1}{4} g_{a\gamma\gamma}F_{\mu\nu}\tilde{F}^{\mu\nu}a, 
\label{eq:lagr}
 \end{equation} 
where $g_{a\gamma\gamma}$ is the coupling constant, $F_{\mu\nu}$ is the photon field strength, 
$ \tilde{F}^{\mu\nu} =  \frac{1}{2} \epsilon^{\mu \nu \alpha \beta} F_{\alpha \beta}$, and $a$ is the axionlike particle field.
For a generic axion, the coupling constant is %$g_{a\gamma\gamma}= \bigl[ 0.203\frac{E}{N}-0.39\bigr] \frac{m_a}{\text GeV^2}$, 
\begin{equation}
g_{a\gamma\gamma}= \bigl[ 0.203\frac{E}{N}-0.39\bigr] \frac{m_a}{\text GeV^2} 
\label{eq:coupl}
\end{equation} 
where $E$ and $N$ are the electromagnetic and color anomalies of the
axial current associated with the axion \cite{cortona,e/n,pdg}. In grand unified models  such as  DFSZ \cite{dfsz} and KSVZ \cite{ksvz}, $E/N = 8/3$ and  $E/N = 0$, respectively,  
while  a broader range of $E/N$ values is possible \cite{e/n,pdg}.  
For the scalar case,  an example of an  $s$ particle weakly coupled to two photons is the dilaton, which arises
in superstring theories  and  interacts with  matter through the trace of the energy-momentum tensor\cite{susy}, and its two-photon interaction
is given by Eq.(\ref{eq:lagr}) with the replacement $\tilde{F}^{\mu\nu} \to F^{\mu\nu}$. 
Usually,  it is assumed that $g_{s\gamma \gamma} = O(M^{-1}_{Pl})$ and that the 
dilaton mass $m_s = O(M_{Pl})$, where $M_{Pl}$ is the Planck mass.\ However, in some models, see, e.g.,  \cite{dilaton}, the  
dilaton could be rather light. Since there are no firm predictions for the coupling $g_{s \gamma \gamma}$
the searches for such particles have become interesting.\   
\par Experimental bounds on $g_{a \gamma \gamma}$ for light $a$'s in the eV-MeV  mass range
 can be obtained  from laser experiments \cite{ruoso, cameron}, from  experiments studying $J/\psi$ and $\Upsilon$ particles \cite{decays}, from the NOMAD experiment by using a photon-regeneration method at the 
CERN SPS neutrino beam \cite{nomad}, and  
from orthopositronium decays \cite{ops}.\ 
  Limits on ALPs in the MeV/$c^2$-GeV/$c^2$ mass range have been typically  placed by beam-dump experiments  or from searches 
 at   $e^+ e^-$ colliders \cite{pdg,babar},  leaving the large area $10^{-2}\lesssim  g_{a\g\g} \lesssim 10^{-5}$ GeV$^{-1}$  of the $(g_{a\g\g}; m_a)$ -parameter space  still unprobed.
 Additionally, since the theory predictions for the coupling, mass scale, and  decay modes of ALPs are still quite uncertain, it is crucial to perform independent
  laboratory tests on the  existence of such particles in the mass  and coupling strength range discussed above.\ 
One possible way to answer these questions is  to search  for ALPs in a beam dump experiment \cite{pdg}.  However, for  the coupling  lying  in the range   
$10^{-2}\lesssim g_{a\g\g} \lesssim 10^{-4}$ GeV$^{-1}$ traditional beam dump experiments are not very promising,  because, for the masses in the sub-GeV/$c^2$ region, the $a$ is expected to be a relatively short-lived particle. 
\par In this Letter, we propose and describe a direct search for  ALPs  with the coupling to  two photons
 from the $(m_a; g_{a\g\g})$ -parameter space uncovered by previous searches.   The  application of the obtained results 
to the  $s\to \g\g$ decay case is straightforward, see e.g.,  \cite{jaec}. 
\par  The  NA64 detector located at the CERN SPS H4 electron beam \cite{h4} is schematically shown in  Fig.~\ref{setup}.
   It consists of a set of 
  beam defining  scintillator counters $S_{1-4}$ and veto  $V_{1,2}$,  a magnetic spectrometer  consisting of two dipole magnets (MBPL1,2)  and a low-material-budget tracker  composed of two upstream Micromegas chambers MM$_{1,2}$,  and four downstream  MM$_{3-6}$  stations \cite{Banerjee:2015eno}, two straw-tube  ST$_{1,2}$ \cite{straw} and GEM$_{1,2}$ chambers . A synchrotron radiation detector (SRD) is used for the identification of incoming $e^-$s \cite{Gninenko:2013rka, na64srd} and suppression of the hadron contamination in the  beam down to the level $\pi/e^- \lesssim 10^{-5}$.  An active dump, consisting of  a preshower detector (PRS) and an electromagnetic (e-m) calorimeter (ECAL),  made of a matrix of $6\times 6 $  Shashlik-type modules,  is  assembled from  Pb and Sc plates of $\simeq 40$ radiation  lengths ($X_0$).  A large  high-efficiency veto counter (VETO) and a massive, hermetic hadronic calorimeter (HCAL) composed of  three modules HCAL1-3 to complete the setup. Each  module is a $3\times3$ cell matrix with a thickness  of $\simeq 7.5$ nuclear interaction lengths.   The events from $e^-$ interactions  in the PRS and ECAL were collected with the  trigger provided by the $S_{1-4}$ 
   requiring also  an in-time cluster in the ECAL with the energy $E_{ECAL} \lesssim 85$ GeV. The detector is described in more detail in Ref.~\cite{na64prd}.
%\section{Search method}
\begin{figure}[tbh!]
\centering
\includegraphics[width=0.4\textwidth]{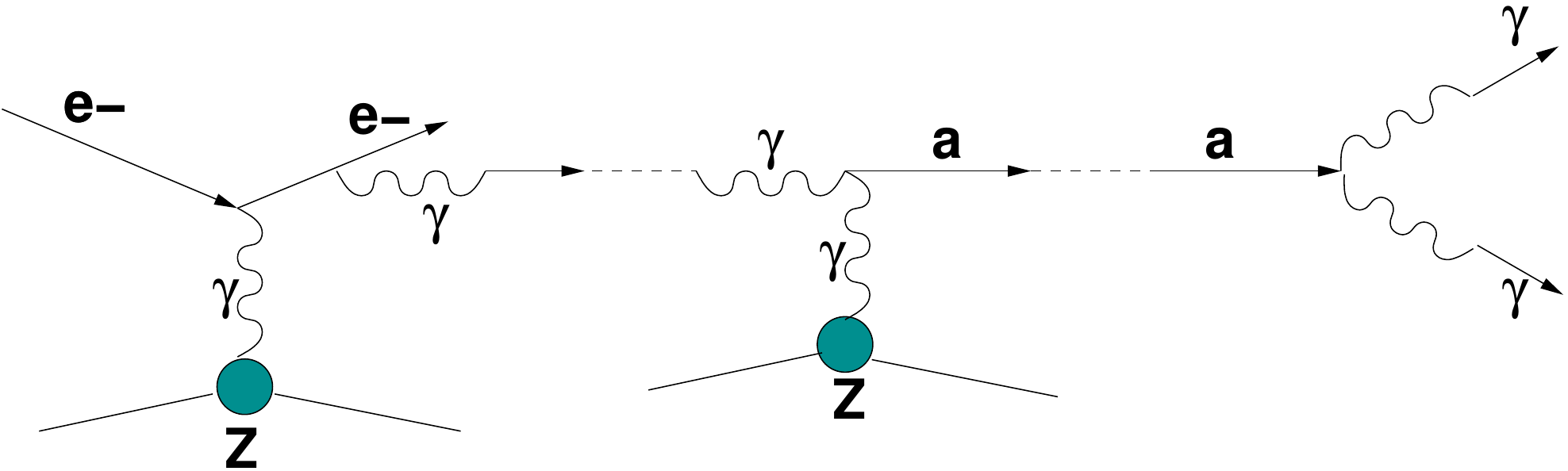}%
\caption{Illustration of the $a$ production and decay in the reaction of Eq.(\ref{eq:prod}).\label{fig:prod}}
\end{figure} 
\par If  ALPs exist, one would expect a  flux of such high energy particles from the dump. 
Both scalars and pseudoscalars could be
produced  in the forward direction through the Primakoff effect 
in interactions of  high energy bremsstrahlung photons,  generated by 100 GeV electrons in the 
target, with virtual photons from  the electrostatic field of the target nuclei:
\begin{equation}
e^-Z \to e^-Z \g; \g Z\to a Z; \agg 
\label{eq:prod}
\end{equation}
as illustrated in Fig.\ref{fig:prod}. 
If the ALP is a relatively  long-lived particle, it  would penetrate the first downstream 
HCAL1 module serving as shielding  and would be observed in the NA64 
detector with two distinctive signatures, 
%referred  to as case 1 and case 2: 
either (1) via  its  decay into $2\g$ inside the HCAL2 or HCAL3 modules (denoted further as HCAL2,3),  or (2) as an event with  large missing energy if it decays downstream of the HCAL2,3. 
 \par  The  selection criteria for signal and background samples have been obtained using a  GEANT4 \cite{Agostinelli:2002hh, geant} based Monte Carlo (MC) simulation of the 
NA64 detector. The code for the simulation of signal events is implemented in the same program according to the
general scheme described in \cite{gkkk, gkkketl}, with the $\agg$ decay width given by $\Gamma_a = g_{a\gamma\gamma}^2 m_a^3/64\pi$.
\begin{figure}[tbh]
\centering
        \includegraphics[width=0.5\textwidth]{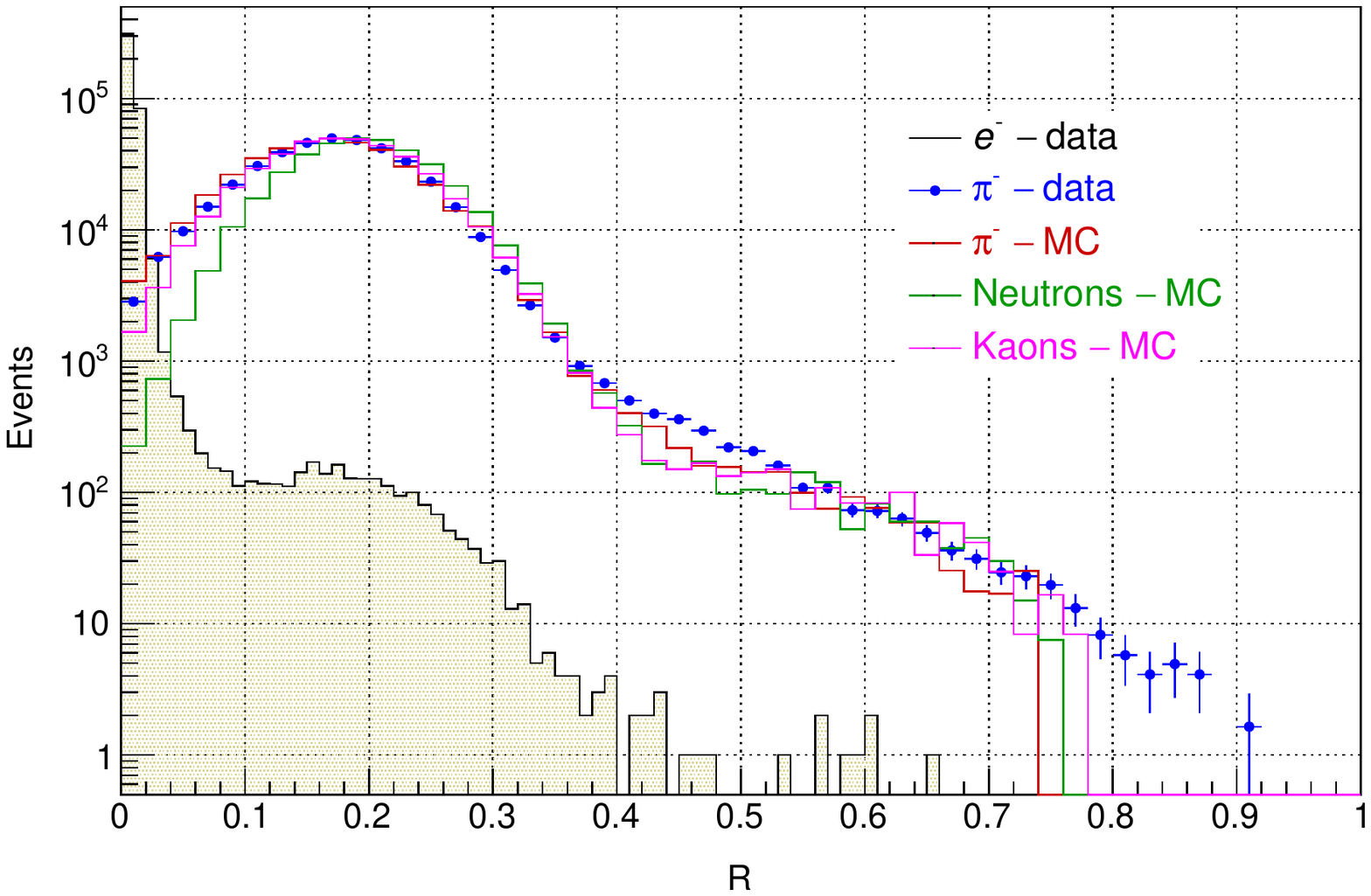}
           \caption{Distributions of the variable $R$ for the 80  GeV $e^-$, $\pi^-, K^0_L$, and $n$   events obtained from data during the ECAL and HCAL calibration runs and simulations. 
%    The electron efficiency and the corresponding fraction of rejected leading hadrons  as a function of $R$ (right panel). 
\label{fig:ratio}}
  \end{figure}
\par The event  from the incoming electron interacting in the dump was required to have 
the incoming  track  momentum in the range of $100\pm 3$ GeV, the SRD signal within the range of synchrotron radiation emitted by $e^-$s, a single PRS cluster  matched to an isolated ECAL 
cluster with an energy greater than 0.5 GeV and an  ECAL cluster with the  shape expected from a single e-m shower \cite{gkkk, na64prd}. 
As the $2\g$ opening angle for the $\agg$ decay is very small, it was not possible to distinguish this decay from a single e-m shower in the HCAL.  Therefore, the candidate events with the signature 1 were selected  as  a single shower in the neutral final state, i.e. no activity in the VETO and the HCAL1, with e-m-like lateral shape,  the shower maximum in the HCAL2,3 central cell and the energy deposition $E_{HCAL}\gtrsim 15$ GeV. This allowed us to reduce background to a small level, while maximizing the $a$ yield by using the cut on the 
ECAL energy $E_{ECAL}\lesssim 85$ GeV. 
For events with  the signature  2, we required the ECAL energy to be $E_{ECAL} \lesssim 50$ GeV and no activity in the VETO and the HCAL.  The above event selection criteria, as well as the efficiency corrections, backgrounds and their systematic errors were similar to those used in our searches for the invisible decays of dark photons \cite{na64prd, na64prl2}.
%Only  $\simeq 1.6 \times 10^4 $ events passed these criteria from combined runs.
\par An additional background suppression for the case 1 was achieved by using the lateral shower shape in the HCAL module. It was  characterized by a variable $R$, defined as $R=\frac{E_{HCAL}-E^{c}_{HCAL}}{E_{HCAL}}$, where $E_{HCAL},~ E^{c}_{HCAL}$ are the total HCAL energy and the energy deposited in the  central cell, respectively.  An example of $R$ distributions  obtained from data and MC simulations is 
 shown in Fig. \ref{fig:ratio}.  As expected,  the distribution for $e^-$s is narrower than for hadrons, and can be  employed  for effective particle identification.  
 Using the cut $R < 0.06$ rejects  $\gtrsim 98\%$ of hadrons, while keeping the signal efficiency    $\gtrsim 95\%$.
\begin{figure*}[tbh!!]
\includegraphics[width=0.33\textwidth]{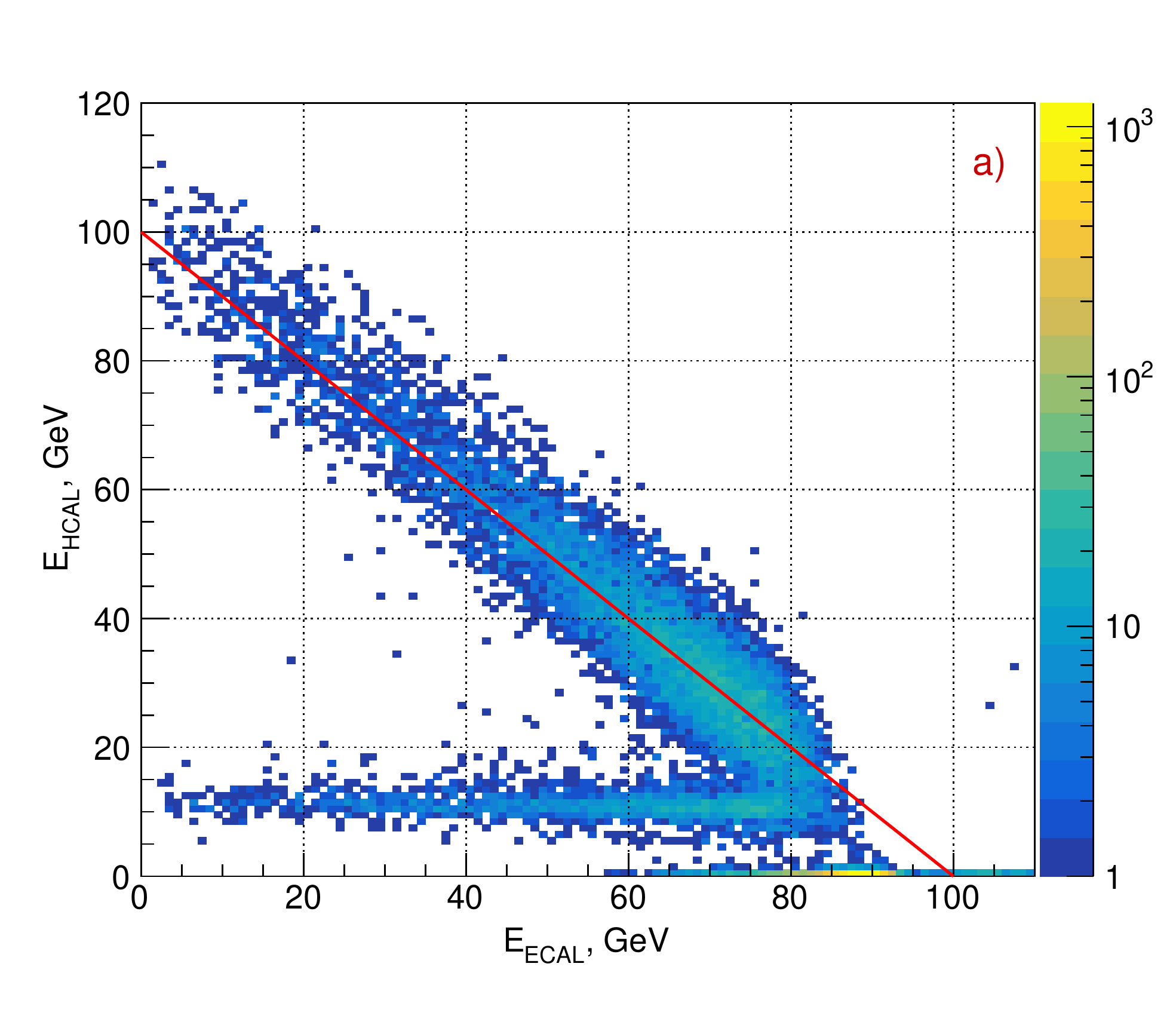}% Here is how to import EPS art
\includegraphics[width=0.33\textwidth]{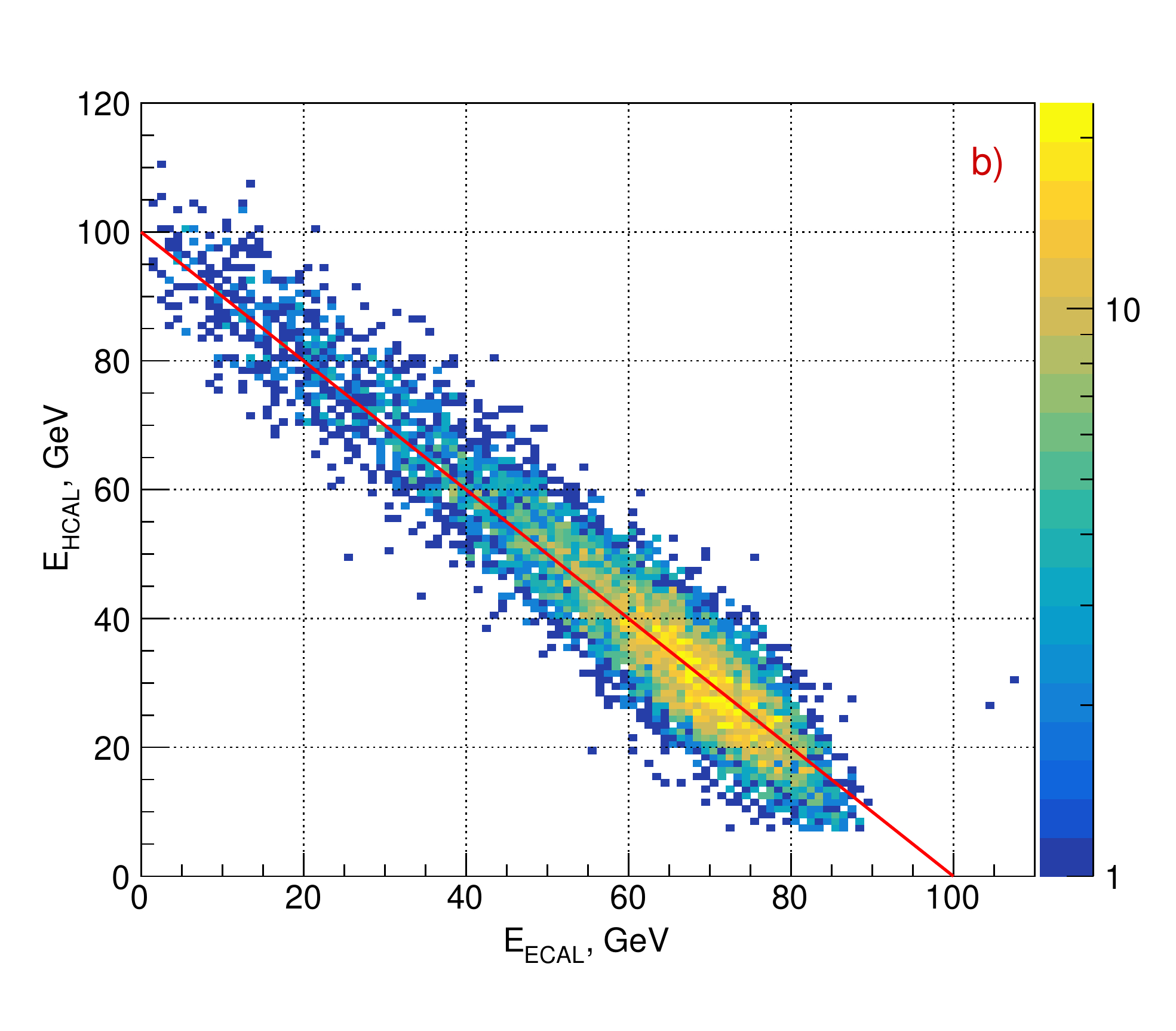}% Here is how to import EPS art
\includegraphics[width=0.33\textwidth]{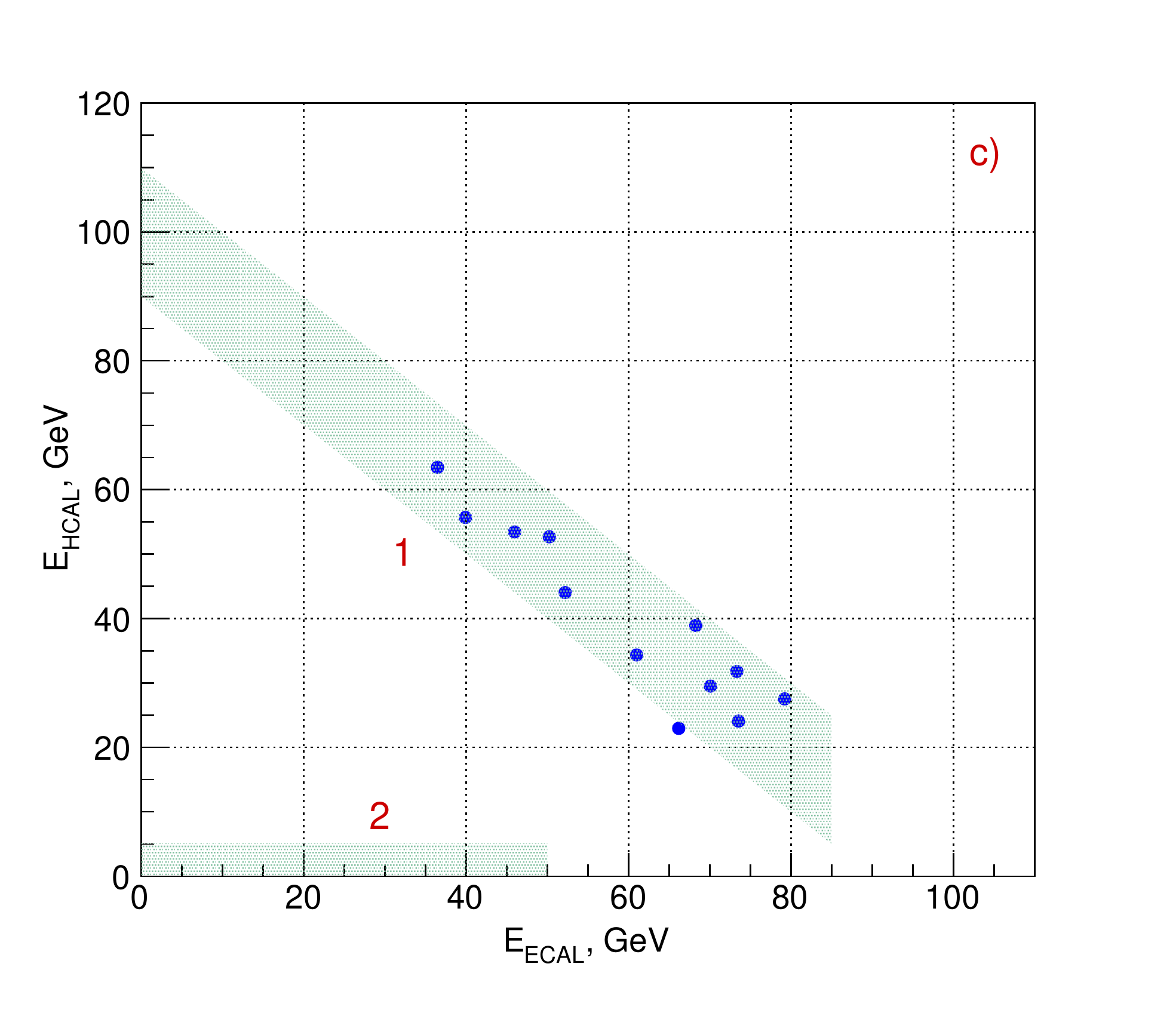}% Here is how to import EPS art
\caption{Panel a) shows the measured  distribution of all events in the ($E_{ECAL}$;$E_{HCAL}$) plane selected at the initial phase of the analysis
with the loose cuts. The distribution of pure neutral hadronic secondaries is illustrated in panel b). 
The shaded area shown in  panel c) represents the signal boxes 1 and 2 in the  $(E_{ECAL};E_{HCAL})$ plane for the signatures 1 and 2 respectively,
where no candidates for the signal events were found  after applying all selection criteria. The blue dots represent 12 events
in the control region $R > 0.06$ from leading neutral hadrons.
The size of the signal box 2 is increased by a factor of 5 along the $E_{HCAL}$ axis for the illustration purposes.
    \label{fig:cand} }
\end{figure*}  
\par The search described in this Letter uses a data samples of $n_{EOT}=2.84\times 10^{11}$  electrons on target (EOT)  
  collected during the 2016-2018 run period  with the beam intensity in the range  $\simeq(2-9)\times 10^6$   e$^-$/spill.
  In Fig.~\ref{fig:cand}a  the distribution of $\simeq 3\times 10^4$ events from the reaction 
$e^- Z \to anything$ in the  $(E_{ECAL}; E_{HCAL})$ plane collected with the trigger and by requiring the presence of a beam 
$e^-$ identified with the SRD tag is shown.
 Events from the horizontal band with $ E_{HCAL}\simeq 10$ GeV originate from the QED dimuon pair production in the ECAL and were  used 
to cross-check  the reliability of the MC simulation  and background estimate \cite{na64prd}.  The further  requirement of no activity in the VETO  identified a sample of  
$\simeq 7\times 10^3$ events  shown in Fig.~\ref{fig:cand}b. This sample corresponds to the neutral  hadronic secondaries  from electroproduction in the dump 
with full hadronic energy deposition in the HCAL1 module. The events located mostly along the diagonal 
  satisfy  the condition of energy conservation $E_{ECAL} + E_{HCAL} \simeq 100$ GeV. 
%The further analysis was performed blindly.
\par  The signal events with the signature 1 are expected to exhibit themselves as an excess of e-m like events 
in the $(E_{ECAL}; E_{HCAL})$ plane in the signal box 1 (Fig.~\ref{fig:cand}c) around the diagonal $E_{ECAL}+E_{HCAL} = 100 \pm 10$ GeV satisfying the energy 
conservation within the energy resolution of the detectors and the cut $R < 0.06$, as shown in Fig.\ref{fig:cand}c. By inverting this cut we obtain the control region, where the  signal events are almost absent.
The signal box 2,   $0 \lesssim E_{ECAL} \lesssim 55$ GeV, $ E_{HCAL} \lesssim 1$ GeV  for signal events  having a large missing energy is also shown \cite{gkkk, gkkketl}. 
\begin{table}[tbh!] 
\begin{center}
\caption{Expected background   for   $2.84\times 10^{11}$ EOT.}\label{tab:bckg}
%\begin{ruledtabular}
\vspace{0.15cm}
\begin{tabular}{lr}
\hline
\hline
Background source& Background, $n_b$\\
\hline
 leading neutrons &$ 0.02\pm 0.008$\\
  leading $K^0$ interactions and  decays &$ 0.14\pm 0.045$\\
beam $\pi,~K $ charge exchange and  decays& $ 0.006\pm 0.002$ \\
dimuons & $<0.001$\\
\hline 
Total $n_b$  &    $0.17\pm 0.046$\\
\hline
\hline 
\end{tabular}
%\end{ruledtabular}
\end{center}
\end{table}
\par The following  processes that may fake the  $\agg$ decay in the HCAL2,3 were considered:
%\begin{enumerate}[(i)]
(i) The production of a leading neutron ($n$),  or  (ii) a leading $K^0$ meson in the ECAL by $e^-$s  in the reaction $e^-A\to n(K^0) + m\pi^0 + X$,  that punchthrough  the HCAL1 and deposited their energy $E_{n(K^0)} \simeq E_0 - E_{ECAL}$  in the HCAL2,3 either in hadronic interactions with a significant e-m component in the shower, or via $\kspio$ or $\klpio$ decays. The reaction can be accompanied by the production of any number $m$ of $\pi^0$s that decay immediately in the  ECAL and a small activity $X$ in the Veto and HCAL1 below of a corresponding thresholds $E_{Veto} \lesssim 0.5~ MIP$ and $E_{HCAL1} \lesssim 1$ GeV. 
(iii) Similar  reactions induced by  beam $\pi^-$ and  $K^-$ that are not rejected by the SRD. As well as
 the $\pi^-,K^-\to e^-\nu $  or $K^- \to \pi^0 e^-\nu  $ decays of  poorly  detected punchthrough beam  $\pi^-,K^-$ downstream of the HCAL1, or 
       production of  a hard bremsstrahlung $\g$ in the downstream part of the HCAL1. (iv) The decays and  reactions induced by muons from dimuon pairs produced  in the ECAL.     
%\end{enumerate}
  \par The main background source is expected from the reactions (ii), mostly due to $K^0_{S,L}$ decays in flight.  
  % To simulate reliably the yield of leading  hadrons is a  difficult, time consuming  task that makes MC with statistics similar to the data not feasible.
 The background was then evaluated by using  the simulation combined with the data themselves by two methods.  In the first one, we use
 the sample of $n_n = 7\times 10^3$ observed  neutral events shown in Fig.\ref{fig:cand}b. A conservative  number of background events originated from leading neutrons and  $K^0$
 was  defined as   $n^{n(K^0)}_b =n_n\times f_{n(K^0)}\times P^{n(K^0)}_{pth} \times P^{n(K^0)}_{em}$, where $f_{n(K^0)}, ~P^{n(K^0)}_{pth}$, and $P^{n(K^0)}_{em}$ are respectively, the fraction of leading neutrons and kaons in the sample, the  probability for  $n (K^0)$ to punchthrough the HCAL1,  and the probability for the $n (K^0)$ induced shower to be accepted as an e-m one. Using GEANT4 simulations we found 
 $f_{n(K^0)}=0.2\pm 0.07(0.18\pm0.06)$. The values  $P^{n(K^0)}_{pth}\simeq 10^{-3} (4.7\times 10^{-3})$ were calculated by using measured absorption cross sections 
 from Refs. \cite{n-inelastic, k-inelastic}.  The values $P^{n(K^0)}_{em}\simeq 5\times10^{-3}(1.1\times10^{-2})$ 
 were evaluated from the MC distributions of Fig. \ref{fig:ratio}.
  The systematic errors of 10\% and 30\% have been assigned to  $P^{n(K^0)}_{pth}$ and  $P^{n(K^0)}_{em}$ values, respectively,   by taking into account the data-MC difference in 
 punchthrough and transverse shapes  of showers (see Fig. \ref{fig:ratio}) generated by $\pi$'s. 
  In the second method we used the number  of $n_c = 12$ neutral events observed in the control region,  shown in Fig.\ref{fig:cand}c. This number was found to be in a 
  good agreement with $9\pm 4$ events expected from the sample of neutral events  shown in Fig.\ref{fig:cand}b. The background then was estimated by
  taking into account the relative composition of these events which was found to be  $\simeq 25\%$ of neutrons and $75\%$ of $K^0$'s. 
\par  All background estimates  were then summed up, taking into account the corresponding normalisation factors. These 
factors were calculated from beam composition,  cross sections for the 
 processes listed above, and punchthrough probabilities  evaluated directly from the data and MC simulations.\   
%The background (iv)  was extracted from the data themselves  by using  the longitudinal segmentation of HCAL for the  conservative punch-through probability estimate. 
The total number  of expected candidate events after applying the selection criteria are given in Table \ref{tab:bckg} for each background component.\ 
The total background of 0.17$\pm$0.046 events, where  statistical and systematic errors were added in quadrature,   estimated with the first method  was found 
to agree with the second estimate resulting in 0.19$\pm$0.07 events.
For the signature 2, the total background in the data sample was estimated to be 0.53$\pm$0.17 events, as described in detail in Ref.\cite{na64prl2}. 
\par After determining all the selection criteria and background levels, we unblinded the signal boxes. No event in the signal boxes shown in
 Fig.\ref{fig:cand}c were found, allowing us to obtain the $m_{a}$-dependent upper limits on the coupling  strength $\gagg$.
 The exclusion limits were calculated by employing the multibin limit setting technique in the
RooStats package \cite{root} with the modified frequentist approach, using  the profile likelihood as a test statistic \cite{junk,limit,Read:2002hq}. 
The combined 90\% confidence level (C.L.) limits on the coupling strength $g_{a\g\g}$ were obtained  from the corresponding 
 limit for the expected number of signal events,  $n_{a}$, which  is given by  the sum: 
\begin{equation}
n_a = \sum_{i=1}^{2} \varepsilon^i_a n^i_a (g_{a\g\g},\ma)
\label{nev}
\end{equation}
where $\varepsilon^i_a$ is the signal efficiency and $n^i_a (g_{a\g\g},\ma)$ is the number of the $a$ decays for the signature {\it i}.
\begin{figure}[tbh]
\begin{center}
\includegraphics[width=0.45\textwidth]{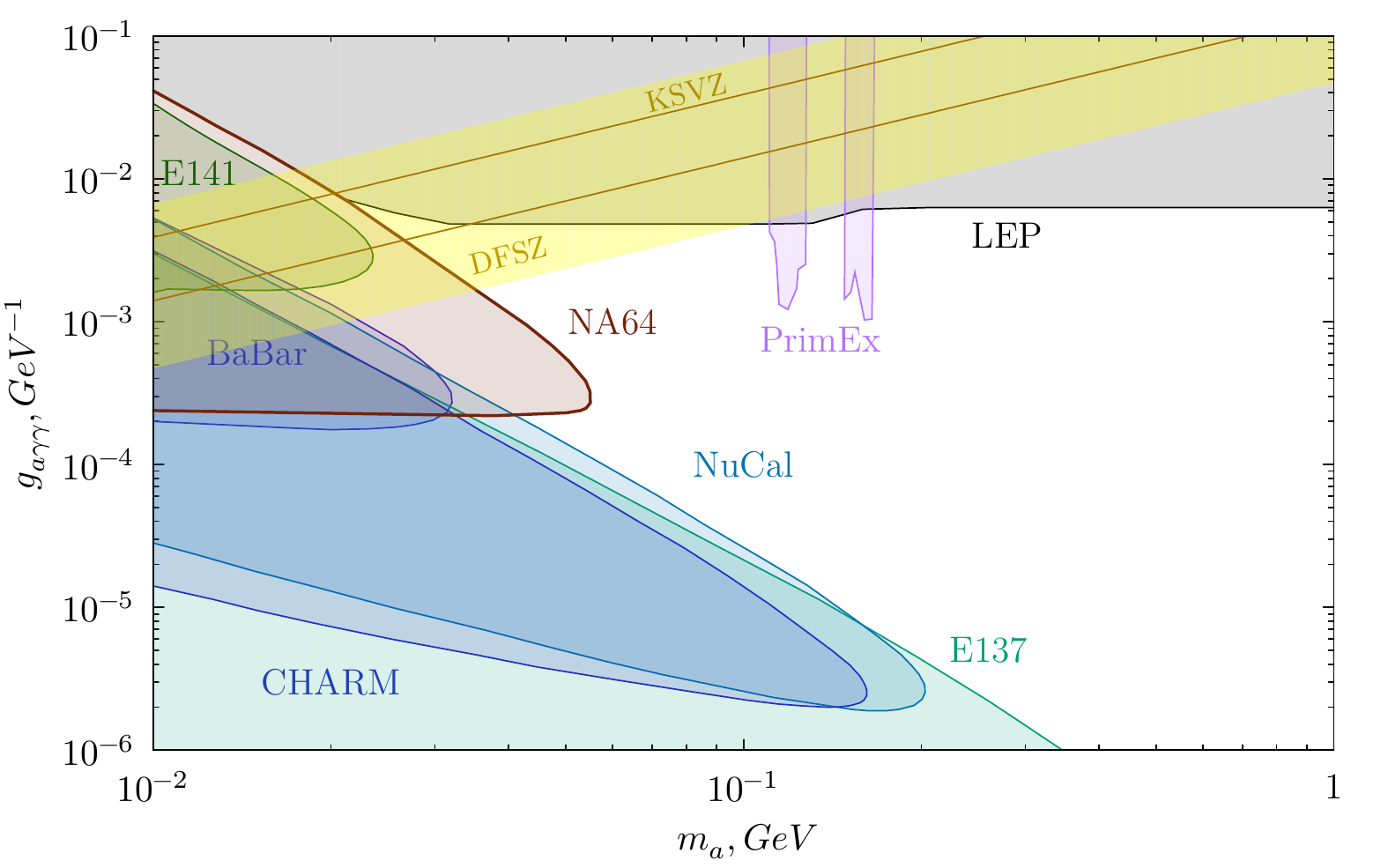}
\caption {The NA64 90\% C.L. exclusion region for ALPs coupling predominantly to photons in the $(m_a ; g_{a\gamma\gamma})$ plane 
 as a function of the (pseudo)scalar mass $m_a$ derived from the present 
analysis.  The yellow band represents the parameter space for the benchmark 
 DFSZ \cite{dfsz} and KSVZ \cite{ksvz} models extended with a broader range of $E/N$ values  \cite{e/n,pdg}. 
 Constraints  from the {\it BABAR} \cite{babar}, E137 \cite{e137}, E141 \cite{e141b, bd},
 LEP \cite{lep},  and PrimEx \cite{primex} experiments, as well as  limits from CHARM \cite{charm} and NuCal \cite{nucal},  updated in Ref.\cite{babette} are also shown. 
   For more limits from indirect searches and  proposed measurements; see, e.g., Refs. \cite{report1,report2, pbc-bsm}. \label{fig:excl}}
\end{center}
\end{figure} 
The $a$ yield from the reaction chain \eqref{eq:prod} was obtained with the calculations described in Ref.\cite{dk} assigning $\lesssim 10\%$ 
 systematic uncertainty  due to different  form-factor parametrizations~\cite{Chen:2011xp,Tsai:1986tx}.
 An additional uncertainty of $\simeq 10\%$ was accounted for the data-MC difference for the dimuon yield \cite{na64prl, na64prd}. 
 The signal detection efficiency for each signature in \eqref{nev} was evaluated by using signal MC and was found slightly $m_a$ dependent.
 For instance, for the signature 1 and $m_a \simeq 10$ MeV, the $\varepsilon^1_a$ and its systematic error was determined from the 
product of efficiencies accounting for the geometrical acceptance ($0.97\pm0.02$), the primary  track ($\simeq 0.83\pm 0.04$), SRD ($\gtrsim 0.95\pm 0.03$), 
ECAL($0.95\pm0.03$), VETO ( $0.94\pm 0.04$), HCAL1 ($0.94\pm 0.04$), and HCAL2,3 ($0.97\pm 0.02$) signal event detection.
 The signal efficiency loss $\lesssim 7\%$ due to pileup  was taken into account using reconstructed dimuon events \cite{na64prd}.
 The VETO and HCAL1 efficiencies were defined as a fraction of events below the corresponding energy thresholds with the main uncertainty estimated
 to be $\lesssim 4\%$ for the signal events, which is caused by the pileup effect from penetrating hadrons.  
% The correction due to the leak of the signal shower energy from the ECAL was simulated for different $a$ masses \cite{gkkk},
% cross-checked with measurements at the $e^- $ beam and found to be small.
 The trigger efficiency was found to be $0.95$ with a small uncertainty of 2\%. 
The total signal efficiency $\epsilon_{a}$ varied from 0.51$\pm$0.09 to 0.48$\pm$0.08 for the $a$ mass range of 10-50 MeV. 
 The total systematic uncertainty on $n_a$ calculated by adding all errors in quadrature did not exceed $20\%$ for both signatures. 
 The attenuation of the $a$ flux due to interactions in the HCAL1 was found to be negligible.
 The combined signal region excluded in the  ($m_a; g_{a\gamma\gamma}$) plane at 90 \% C.L. 
is shown in Fig.~\ref{fig:excl} together with the results of other experiments. Our limits are valid for both scalar and pseudoscalar cases and 
exclude the region in the coupling range $2\times  10^{-4} \lesssim g_{a\g\g} \lesssim 5 \times 10^{-2}$ GeV$^{-1}$ for masses $\ma \lesssim 55$ MeV. 
  \par 
% This work  serves as a memorial to Danila Tlisov and his contributions to the NA64 collaboration.
We gratefully acknowledge the support of the CERN management and staff 
and the technical staff of the participating institutions for their vital contributions. We would like to thank M.W.~Krasny for providing us  with Ref. cite{e141b} and useful comments,  B.~D\"obrich for providing information  on the E141 and updated CHARM and NuCal  exclusion curves, and   G.~Lanfranchi for  valuable discussions.   
 This work was supported by the  Helmholtz-Institut f\"ur Strahlen- und Kern-physik (HISKP), University of Bonn, the Carl Zeiss Foundation Grant No. 0653-2.8/581/2, and  Verbundprojekt-05A17VTA-CRESST-XENON (Germany), Joint Institute for Nuclear Research (JINR) (Dubna),   the Ministry of Science and Higher Education (MSHE) in the frame of the Agreement No. 05.613.21.0098 ID No. RFMEFI61320X0098 on July 23, 2020, TPU Competitiveness Enhancement Program and RAS (Russia),
 ETH Zurich and SNSF Grants No. 169133, 186181, and No. 186158 (Switzerland), and FONDECYT Grants  No. 1191103, No. 190845, and No. 3170852, UTFSM PI~M~18~13, ANID PIA/APOYO AFB180002 (Chile).

\end{document}

%% file: author_list.tex
% remove these 3 lines before journal submittal.
%\centerline{Preliminary author list dated 20 September  2016}
% end removal before journal submittal
\affiliation{\it Universit\"at Bonn, Helmholtz-Institut f\"ur Strahlen-und Kernphysik, 53115 Bonn, Germany} 
\affiliation{\it Joint Institute for Nuclear Research, 141980 Dubna, Russia}
\affiliation{\it Technische Universit\"at M\"unchen, Physik  Department, 85748 Garching, Germany}
\affiliation{\it CERN, European Organization for Nuclear Research, CH-1211 Geneva, Switzerland}
\affiliation{\it University of Illinois at Urbana Champaign, Urbana, 61801-3080 Illinois, USA}
\affiliation{\it UCL Departement of Physics and Astronomy, University College London, Gower St. London WC1E 6BT, United Kingdom}
\affiliation{\it Institute for Nuclear Research, 117312 Moscow, Russia}
\affiliation{\it P.N. Lebedev Physical Institute, Moscow, Russia, 119 991 Moscow, Russia}
\affiliation{\it Skobeltsyn Institute of Nuclear Physics, Lomonosov Moscow State University, 119991  Moscow, Russia}
\affiliation{\it Physics Department, University of Patras, 265 04 Patras, Greece} 
\affiliation{\it State Scientific Center of the Russian Federation Institute for High Energy Physics of National Research Center 'Kurchatov Institute' (IHEP), 142281 Protvino, Russia}
\affiliation{\it Departamento de Ciencias F\'{i}sicas, Universidad Andres Bello, Sazi\'{e} 2212, Piso 7, Santiago, Chile}
\affiliation{\it Tomsk Polytechnic University, 634050 Tomsk, Russia}
\affiliation{\it Tomsk State Pedagogical University, 634061 Tomsk, Russia}
\affiliation{\it Universidad T\'{e}cnica Federico Santa Mar\'{i}a, 2390123 Valpara\'{i}so, Chile}
\affiliation{\it ETH Z\"urich, Institute for Particle Physics and Astrophysics, CH-8093 Z\"urich, Switzerland}
\author{D.~Banerjee}\affiliation{\it CERN, European Organization for Nuclear Research, CH-1211 Geneva, Switzerland}\affiliation{\it University of Illinois at Urbana Champaign, Urbana, 61801-3080 Illinois, USA}
\author{J.~Bernhard}\affiliation{\it CERN, European Organization for Nuclear Research, CH-1211 Geneva, Switzerland}
\author{V.~E.~Burtsev}\affiliation{\it Joint Institute for Nuclear Research, 141980 Dubna, Russia}
\author{A.~G.~Chumakov}\affiliation{\it Tomsk State Pedagogical University, 634061 Tomsk, Russia}
\author{D.~Cooke}\affiliation{\it UCL Departement of Physics and Astronomy, University College London, Gower St. London WC1E 6BT, United Kingdom}
%\author{P.~Crivelli\footnote[1]{Corresponding author, Paolo.Crivelli@cern.ch}}\affiliation{\it ETH Z\"urich, Institute for Particle Physics and Astrophysics, CH-8093 Z\"urich, Switzerland}
\author{P.~Crivelli\footnote[1]{Corresponding author: Paolo.Crivelli@cern.ch}}
%\email[Corresponding author: Paolo.Crivelli@cern.ch, ]{Sergei.Gninenko@cern.ch}
\affiliation{\it ETH Z\"urich, Institute for Particle Physics and Astrophysics, CH-8093 Z\"urich, Switzerland}
\author{E.~Depero}\affiliation{\it ETH Z\"urich, Institute for Particle Physics and Astrophysics, CH-8093 Z\"urich, Switzerland}
\author{A.~V.~Dermenev}\affiliation{\it Institute for Nuclear Research, 117312 Moscow, Russia}
\author{S.~V.~Donskov}\affiliation{\it State Scientific Center of the Russian Federation Institute for High Energy Physics of National Research Center 'Kurchatov Institute' (IHEP), 142281 Protvino, Russia}
%\author{F.~Dubinin}\affiliation{\it P.N. Lebedev Physics Institute, Moscow, Russia, 119 991 Moscow, Russia}
\author{R.~R.~Dusaev}\affiliation{\it Tomsk Polytechnic University, 634050 Tomsk, Russia}
%\author{S.~Emmenegger}\affiliation{\it ETH Z\"urich, Institute for Particle Physics, CH-8093 Z\"urich, Switzerland}
\author{T.~Enik}\affiliation{\it  Joint Institute for Nuclear Research, 141980 Dubna, Russia}
\author{N.~Charitonidis}\affiliation{\it CERN, European Organization for Nuclear Research, CH-1211 Geneva, Switzerland}
\author{A.~Feshchenko}\affiliation{\it  Joint Institute for Nuclear Research, 141980 Dubna, Russia}
\author{V.~N.~Frolov}\affiliation{\it  Joint Institute for Nuclear Research, 141980 Dubna, Russia}
\author{A.~Gardikiotis}\affiliation{\it Physics Department, University of Patras, 265 04 Patras, Greece}
\author{S.~G.~Gerassimov }\affiliation{\it Technische Universit\"at M\"unchen, Physik  Department, 85748 Garching, Germany}\affiliation{\it P.N. Lebedev Physical Institute, Moscow, Russia, 119 991 Moscow, Russia}
\author{S.~N.~Gninenko\footnote[1]{Corresponding author: Sergei.Gninenko@cern.ch}}
%\thanks{ Sergei.Gninenko@cern.ch}
\affiliation{\it Institute for Nuclear Research, 117312 Moscow, Russia}
%\author{S.~N.~Gninenko\footnote[1]{Corresponding author: Sergei.Gninenko@cern.ch}}\affiliation{\it Institute for Nuclear Research, 117312 Moscow, Russia}
\author{M.~H\"osgen}\affiliation{\it Universit\"at Bonn, Helmholtz-Institut f\"ur Strahlen-und Kernphysik, 53115 Bonn, Germany}
\author{M.~Jeckel}\affiliation{\it CERN, European Organization for Nuclear Research, CH-1211 Geneva, Switzerland}
\author{V.~A.~Kachanov}\affiliation{\it State Scientific Center of the Russian Federation Institute for High Energy Physics of National Research Center 'Kurchatov Institute' (IHEP), 142281 Protvino, Russia}
\author{A.~E.~Karneyeu}\affiliation{\it Institute for Nuclear Research, 117312 Moscow, Russia}
\author{G.~Kekelidze}\affiliation{\it  Joint Institute for Nuclear Research, 141980 Dubna, Russia}
\author{B.~Ketzer}\affiliation{\it Universit\"at Bonn, Helmholtz-Institut f\"ur Strahlen-und Kernphysik, 53115 Bonn, Germany}
\author{D.~V.~Kirpichnikov}\affiliation{\it Institute for Nuclear Research, 117312 Moscow, Russia}
\author{M.~M.~Kirsanov}\affiliation{\it Institute for Nuclear Research, 117312 Moscow, Russia}
\author{V.~N.~Kolosov}\affiliation{\it State Scientific Center of the Russian Federation Institute for High Energy Physics of National Research Center 'Kurchatov Institute' (IHEP), 142281 Protvino, Russia}
\author{I.~V.~Konorov}\affiliation{\it Technische Universit\"at M\"unchen, Physik  Department, 85748 Garching, Germany}\affiliation{\it P.N. Lebedev Physical Institute, Moscow, Russia, 119 991 Moscow, Russia} 
\author{S.~G.~Kovalenko}\affiliation{\it Departamento de Ciencias F\'{i}sicas, Universidad Andres Bello, Sazi\'{e} 2212, Piso 7, Santiago, Chile}
\author{V.~A.~Kramarenko}\affiliation{\it  Joint Institute for Nuclear Research, 141980 Dubna, Russia}\affiliation{\it Skobeltsyn Institute of Nuclear Physics, Lomonosov Moscow State University, 119991  Moscow, Russia}
\author{L.~V.~Kravchuk}\affiliation{\it Institute for Nuclear Research, 117312 Moscow, Russia}
\author{ N.~V.~Krasnikov}\affiliation{\it  Joint Institute for Nuclear Research, 141980 Dubna, Russia}\affiliation{\it Institute for Nuclear Research, 117312 Moscow, Russia}
\author{S.~V.~Kuleshov}\affiliation{\it Departamento de Ciencias F\'{i}sicas, Universidad Andres Bello, Sazi\'{e} 2212, Piso 7, Santiago, Chile}
\author{V.~E.~Lyubovitskij}\affiliation{\it Tomsk State Pedagogical University, 634061 Tomsk, Russia}\affiliation{\it Universidad T\'{e}cnica Federico Santa Mar\'{i}a, 2390123 Valpara\'{i}so, Chile}
\author{V.~Lysan}\affiliation{\it  Joint Institute for Nuclear Research, 141980 Dubna, Russia}
\author{V.~A.~Matveev}\affiliation{\it  Joint Institute for Nuclear Research, 141980 Dubna, Russia}
\author{Yu.~V.~Mikhailov}\affiliation{\it State Scientific Center of the Russian Federation Institute for High Energy Physics of National Research Center 'Kurchatov Institute' (IHEP), 142281 Protvino, Russia}
\author{L.~Molina Bueno}\affiliation{\it ETH Z\"urich, Institute for Particle Physics and Astrophysics, CH-8093 Z\"urich, Switzerland}
%\author{V.~V.~Myalkovskiy}\affiliation{\it  Joint Institute for Nuclear Research, 141980 Dubna, Russia}
%\author{V.~D.~Peshekhonov\footnote{Deceased}}\affiliation{\it  Joint Institute for Nuclear Research, 141980 Dubna, Russia}
\author{D.~V.~Peshekhonov}\affiliation{\it  Joint Institute for Nuclear Research, 141980 Dubna, Russia}
%\author{O.~Petuhov}\affiliation{\it Institute for Nuclear Research, 117312 Moscow, Russia} 
\author{V.~A.~Polyakov}\affiliation{\it State Scientific Center of the Russian Federation Institute for High Energy Physics of National Research Center 'Kurchatov Institute' (IHEP), 142281 Protvino, Russia}
\author{B.~Radics}\affiliation{\it ETH Z\"urich, Institute for Particle Physics and Astrophysics, CH-8093 Z\"urich, Switzerland}
\author{R.~Rojas}\affiliation{\it Universidad T\'{e}cnica Federico Santa Mar\'{i}a, 2390123 Valpara\'{i}so, Chile}
\author{A.~Rubbia}\affiliation{\it ETH Z\"urich, Institute for Particle Physics and Astrophysics, CH-8093 Z\"urich, Switzerland}
\author{V.~D.~Samoylenko}\affiliation{\it State Scientific Center of the Russian Federation Institute for High Energy Physics of National Research Center 'Kurchatov Institute' (IHEP), 142281 Protvino, Russia}
\author{H.~Sieber}\affiliation{\it ETH Z\"urich, Institute for Particle Physics and Astrophysics, CH-8093 Z\"urich, Switzerland}
\author{D.~Shchukin}\affiliation{\it P.N. Lebedev Physical Institute, Moscow, Russia, 119 991 Moscow, Russia}
\author{V.~O.~Tikhomirov}\affiliation{\it P.N. Lebedev Physical Institute, Moscow, Russia, 119 991 Moscow, Russia}
\author{I.~Tlisova}\affiliation{\it Institute for Nuclear Research, 117312 Moscow, Russia} 
\author{D.~A.~Tlisov\footnote[2]{Deceased}}\affiliation{\it Institute for Nuclear Research, 117312 Moscow, Russia} 
\author{A.~N.~Toropin}\affiliation{\it Institute for Nuclear Research, 117312 Moscow, Russia}
\author{A.~Yu.~Trifonov}\affiliation{\it Tomsk State Pedagogical University, 634061 Tomsk, Russia}
\author{B.~I.~Vasilishin}\affiliation{\it Tomsk Polytechnic  University, 634050 Tomsk, Russia}
\author{G.~Vasquez Arenas}\affiliation{\it Universidad T\'{e}cnica Federico Santa Mar\'{i}a, 2390123 Valpara\'{i}so, Chile}
\author{P.~V.~Volkov}\affiliation{\it  Joint Institute for Nuclear Research, 141980 Dubna, Russia}\affiliation{\it Skobeltsyn Institute of Nuclear Physics, Lomonosov Moscow State University, 119991  Moscow, Russia}
\author{V.~Yu.~Volkov}\affiliation{\it Skobeltsyn Institute of Nuclear Physics, Lomonosov Moscow State University, 119991  Moscow, Russia}
\author{P.~Ulloa}\affiliation{\it Departamento de Ciencias F\'{i}sicas, Universidad Andres Bello, Sazi\'{e} 2212, Piso 7, Santiago, Chile}
%\author{K.~Zhukov}\affiliation{\it P.N. Lebedev Physics Institute, Moscow, Russia, 119 991 Moscow, Russia}
%\author{K.~Zioutas}\affiliation{\it Physics Department, University of Patras, Patras, Greece} 

%
% visitor_addresses.tex                       19 August 2015
%  available symbols are:
%  $\ast, \dag, \ddag, \S, \P, $\|$, $\ast\ast$, \dag\dag, \ddag\ddag ,\#
%
%\collaboration{The NA64 Collaboration\footnote{https://na64.web.cern.ch}}\noaffiliation
\collaboration{The NA64 Collaboration}\noaffiliation
\vskip 0.25cm

%% file: alp_arxiv_rev.bbl
\begin{thebibliography}{99}
\bibitem{ww}
S.~Weinberg, Phys. Rev. Lett. {\bf 40}, 223 (1978);\\
F.~Wilczek, Phys. Rev. Lett. {\bf 40}, 279 (1978).

\bibitem{pq}R. D.~Peccei and H. R.~Quinn, Phys. Rev. Lett. {\bf 38}, 1440 (1977).

\bibitem{dfsz} M.~Dine, W.~Fischler, and M.~Srednicki, 
Phys. Lett. {\bf B104}, 199 (1981);\\
A.~Zhitnitski, Sov. J. Nucl. Phys. {\bf 31}, 260 (1980) 
[Yad. Fiz. {\bf 31}, 497 (1980).

\bibitem{ksvz} J.~E.~Kim, Phys. Rev. Lett. {\bf 43}, 103 (1979);\\
M.~Schifman, A.~Vainstein and V.~Zakharov, Nucl. Phys. {\bf 166}, 493 (1981).

\bibitem{review}
N.~V.~Krasnikov, V.~A.~Matveev and A.N.~Tavkhelidze, 
Sov. J. Part. Nucl. {\bf 12}, 38 (1981);\\
J.~E.~Kim, Phys. Rep. {\bf 150}, 1 (1987);\\
H.~Y.~ Cheng, Phys. Rep. {\bf 158}, 1 (1988);\\
G.~Raffelt, Phys. Rep. {\bf 198}, 1 (1990).
\bibitem{cortona} G.~G.~di Cortona et al., JHEP {\bf 1601}, 034 (2016).

\bibitem{e/n} J.E. Kim, Phys. Rev. D {\bf 58}, 055006 (1998);
L. Di Luzio, F. Mescia and E. Nardi, Phys. Rev. Lett. {\bf 118}, 031801 (2017).

\bibitem{pdg} 
M.~Tanabashi {\it et al.}  (Particle Data Group), 
Phys. Rev. D {\bf 98}, 030001 (2018).
  
\bibitem{g-2e}
R.~H.~Parker, C.~Yu.~W.~Zhong, B.~Estey, and H.~M\"uller, 
Science {\bf 360}, 191 (2018).

\bibitem{g-2mu}
G.~W.~Bennett {\it et al.}  (Muon g-2 Collaboration), 
Phys. Rev. D {\bf 73}, 072003 (2006).

\bibitem{marci}
H.~Davoudiasl and W.~J.~Marciano, Phys. Rev. D {\bf 98}, 075011 (2018); 
C.-Yi~Chen, H.~Davoudiasl, W.~J.~Marciano, and C.~Zhang, 
Phys. Rev. D {\bf 93}, 036006 (2016);
W.~J.~Marciano, A.~Masiero, P.~Paradisi, and M.~Passera, 
Phys. Rev. D {\bf 94}, 115033 (2016);
F.~Abu-Ajamieh, arXiv:1810.08891. 

\bibitem{Essig:2013lka} 
  R.~Essig {\it et al.}, arXiv:1311.0029.

\bibitem{report1}
  J.~Alexander {\it et al.}, arXiv:1608.08632.

\bibitem{report2} 
  M.~Battaglieri {\it et al.}, arXiv:1707.04591.

\bibitem{pbc-bsm}
J.~Beacham  {\it et al.}, J. Phys. G {\bf 47}, 010501 (2020).   

\bibitem{pbc}
R.~Alemany  {\it et al.}, 
 arXiv:1902.00260.

\bibitem{berlin}
A.~Berlin, N.~Blinov, G.~Krnjaic, P.~Schuster, and N.~Toro,
Phys. Rev. D {\bf 99}, 075001 (2019).

\bibitem{Feng:2018noy}
  J.~L.~Feng, I.~Galon, F.~Kling, and S.~Trojanowski,
  %``Axionlike particles at FASER: The LHC as a photon beam dump,''
  Phys.\ Rev.\ D {\bf 98}, 055021 (2018). 

\bibitem{Dobrich:2015jyk}
  B.~D\"obrich, J.~Jaeckel, F.~Kahlhoefer, A.~Ringwald, 
 and K.~Schmidt-Hoberg,
  %``ALPtraum: ALP production in proton beam dump experiments,''
  JHEP {\bf 1602}, 018 (2016) 018.  

\bibitem{Bauer:2018uxu}
  M.~Bauer, M.~Heiles, M.~Neubert, and A.~Thamm,
  %``axionlike Particles at Future Colliders,''
  Eur.\ Phys.\ J.\ C {\bf 79}, 74 (2019). 

\bibitem{jaec}
 J.~Jaeckel and M.~Spannowsky,  Phys. Lett. B {\bf 753}, 482 (2016).

\bibitem{gkm}	
S.~N.~Gninenko, N.~V.~Krasnikov, and V.~A.~Matveev, arXiv:2003.07257.


\bibitem{dvk}
    D.~V.~Kirpichnikov, V.~ E.~Lyubovitskij, A.~S.~Zhevlakov,  arXiv: 2002.07496. 

\bibitem{susy}
P.~Horava and E.~Witten, Nucl. Phys. {\bf B460}, 506 (1996);  {\bf B475}, 94 (1996).

\bibitem{dilaton}
N.~Arkani-Hamed, S.~Dimopoulos, and G.~Dvali, 
Phys. Lett. {\bf B429}, 263 (1998);\\
I.~Antoniadis {\it et al.},  
Phys. Lett. {\bf B436}, 257 (1998). 

\bibitem{ruoso} G.~Ruoso {\it et al.}, 
Z. Phys. {\bf C56}, 505 (1992). 

\bibitem{cameron} R.~Cameron {\it et al.}, 
Phys. Rev. {\bf D47}, 3707 (1993). 

\bibitem{decays} M.~S.~Alam {\it et al.}, 
Phys. Rev. {\bf D27}, 1665 (1983); 
N.~J.~Baker {\it et al.}, 
Phys. Rev. Lett. {\bf 59}, 2832 (1987). 

\bibitem{nomad}
P.~Astier {\it et al.}, (NOMAD Collaboration), 
Phys.\ Lett.\  B {\bf 479}, 371 (2000). 

\bibitem{ops}
U.~Amaldi, G.~Carboni, B.~Jonson, and J.~Thun, 
Phys. Lett. {\bf B153}, 444 (1985);\\
S.~Orito {\it et al.}, Phys. Rev. Lett. {\bf 63}, 597 (1989);\\
M.~V.~Akopian, G.~S.~Atoyan, S.N.~Gninenko, and V.~V.~Sukhov, 
Phys. Lett. {\bf B272}, 443 (1991);\\
S.~N.~Gninenko, Yu.~M.~Klubakov, A.~A.~Poblaguev, and V.~E.~Postoev, 
Phys. Lett. {\bf B237}, 287 (1990);\\
T.~Maeno {\it et al.}, Phys. Lett. {\bf B 351}, 574 (1995);\\
S.~Asai, S.~Orito, K.~Yoshimura, and T.~Haga,  
Phys. Rev. Lett. {\bf 66}, 2440 (1991).

\bibitem{babar}
M.~J.~Dolan, T.~Ferber, C.~Hearty, F.~Kahlhoefer, and 
K.~Schmidt-Hoberg, JHEP {\bf 12}, 094 (2017).

\bibitem{h4} See, for example, http://sba.web.cern.ch/sba/

\bibitem{Banerjee:2015eno}
  D.~Banerjee, P.~Crivelli, and A.~Rubbia,
   Adv.\ High Energy Phys.\  {\bf 2015}, 105730 (2015).

\bibitem{straw}  V.~Yu.~Volkov, P.~V.~Volkov, T.~L.~Enik, G.~D.~Kekelidze, V.~A.~Kramarenko, V.~M.~Lysan, D.~V.~Peshekhonov, A.~A.~Solin,  A.~V.~Solin, Phys. Part. Nucl. Lett. {\bf 16}, 847 (2019).
  
\bibitem{Gninenko:2013rka}
S.~N.~Gninenko,
  Phys.\ Rev.\ D {\bf 89}, 075008 (2014).

\bibitem{na64srd}  
  E.~Depero {\it et al.},
  Nucl.\ Instrum.\ Methods\ Phys. Res., Sect. A {\bf 866}, 196 (2017).

\bibitem{na64prd} 
  D.~Banerjee {\it et al.} (NA64 Collaboration),
  Phys.\ Rev.\ D {\bf 97},  072002 (2018).
  
\bibitem{Agostinelli:2002hh}
  S.~Agostinelli {\it et al.} (GEANT4 Collaboration),
  Nucl.\ Instrum.\ Methods\ Phys. Res., Sect. A {\bf 506}, 250 (2003).

\bibitem{geant} 
  J.~Allison {\it et al.},
  IEEE Trans.\ Nucl.\ Sci.\  {\bf 53}, 270 (2006).

\bibitem{na64prl2} 
  D.~Banerjee {\it et al.} (NA64 Collaboration),
  Phys.\ Rev.\ Lett. {\bf 123},  121801 (2019).

\bibitem{gkkk} 
  S.~N.~Gninenko, N.~V.~Krasnikov, M.~M.~Kirsanov, and D.~V.~Kirpichnikov,
  Phys.\ Rev.\ D {\bf 94}, 095025 (2016).
  
\bibitem{gkkketl}
  S.~N.~Gninenko, D.~V.~Kirpichnikov, M.~M.~Kirsanov, and N.~V.~Krasnikov,   
  Phys. Lett. B {\bf 782}, 406 (2018).

 \bibitem{n-inelastic}
  T.J. Roberts, H.R. Gustafson, L.W. Jones, M.J, Longo and M.R. Whalley, 
Nucl. Phys. B {\bf 159}, 56 (1979). 

 \bibitem{k-inelastic}
 A.S. Carroll {\it et al.}, Phys. Lett B {\bf 80B}, 319 (1979). 


\bibitem{na64be} 
  D.~Banerjee {\it et al.} (NA64 Collaboration),
  Phys. Rev. Lett. {\bf 120},  231802 (2018).

\bibitem{root} 
  I.~Antcheva {\it et al.},
  Comput.\ Phys.\ Commun.\  {\bf 180}, 2499 (2009).

\bibitem{junk}
T.~Junk, 
Nucl. Instrum. Methods Phys. Res., Sect. A {\bf 434}, 435 (1999).

\bibitem{limit}
G.~Cowan, K.~Cranmer, E.~Gross, and O.~Vitells, 
Eur. Phys. J. C {\bf 71}, 1  (2011).

\bibitem{Read:2002hq}
  A.~L.~Read,
  J.\ Phys.\ G {\bf 28}, 2693 (2002).

\bibitem{dk} 
D.~V.~Kirpichnikov, R.~R.~Dusaev, and M.~M.~Kirsanov, 
%``Photoproduction of axionlike particles at NA64,''
  arXiv:2004.04469.
  

%\cite{Chen:2011xp}
\bibitem{Chen:2011xp}
  Y.~Z.~Chen, Y.~A.~Luo, L.~Li, H.~Shen and X.~Q.~Li,
  %``Determining the nuclear form factor for detection of dark matter in the relativistic mean field theory,''
  Commun.\ Theor.\ Phys.\  {\bf 55} (2011) 1059
 % doi:10.1088/0253-6102/55/6/21
  [arXiv:1101.3049 [hep-ph]].
  %%CITATION = doi:10.1088/0253-6102/55/6/21;%%
  %8 citations counted in INSPIRE as of 11 Apr 2020  
  
%\cite{Tsai:1986tx}
\bibitem{Tsai:1986tx}
Y.~Tsai,
%``AXION BREMSSTRAHLUNG BY AN ELECTRON BEAM,''
Phys.\ Rev.\ D \textbf{34} (1986), 1326
%doi:10.1103/PhysRevD.34.1326
%56 citations counted in INSPIRE as of 12 Apr 2020  

 \bibitem{na64prl} 
  D.~Banerjee {\it et al.} (NA64 Collaboration),
  Phys.\ Rev.\ Lett.\  {\bf 118}, 011802 (2017).


\bibitem{e137}
J. D. Bjorken  {\it et al.}, 
Phys. Rev. D {\bf 38}, 3375 (1988).

\bibitem{e141b}
M.~W.~Krasny {\it et al.} (E141 Collaboration), 
%"Recent searches for pseudoscalar bosons in electron beam-dump experiments", 
Preprint Univ. of  Rochester, UR-1029 (1987); 
The E141 limits in $(\gagg; m_a)$ plane were obtained in Ref.\cite{bd}.

\bibitem{bd}
B.~D\"obrich, 
%"axionlike Particles from Primakov production in beam-dumps"  
CERN Proc.\ {\bf 1}, 253 (2018). 

\bibitem{lep}
G.~Abbiendi {\it et al.} (OPAL Collaboration), 
Eur. Phys. J. C {\bf 26}, 331 (2003).

\bibitem{primex}
D.~Aloni, C.~Fanelli, Y.~Soreq, and M.~Williams, 
Phys.\ Rev.\  Lett.  {\bf 123},  071801 (2019).

\bibitem{charm}
F.~Bergsma {\it et al.}, (CHARM Collaboration), 
Phys.\ Lett.\  B {\bf 157}, 458 (1985).

\bibitem{nucal}
J. Bl\"umlein {\it et al.}, 
Z. Phys. C {\bf 51}, 341 (1991).

\bibitem{babette}
B.~D\"obrich, J.~Jaeckel and T.~Spadaro,
%``Light in the beam dump. axionlike Particle production from decay photons in proton beam-dumps,''
JHEP \textbf{05}, 213 (2019)
%doi:10.1007/JHEP05(2019)213
%[arXiv:1904.02091 [hep-ph]].


\end{thebibliography}
